%% file: main.tex
\documentclass[conference]{IEEEtran}
\IEEEoverridecommandlockouts
\usepackage{cite}
\usepackage{hyperref}
\usepackage{amsmath,amssymb,amsfonts}
\usepackage{algorithmic}
\usepackage{graphicx}
\usepackage{textcomp}
\usepackage{xcolor}
\usepackage{float}
\usepackage{graphicx}
\usepackage{subcaption}
\usepackage{booktabs}
\usepackage[ruled, lined,linesnumbered, longend]{algorithm2e}
\definecolor{myred}{RGB}{255,230,230}
\definecolor{myblue}{RGB}{222,235,247}
\definecolor{mygreen}{RGB}{226,240,217}
\definecolor{myyellow}{RGB}{255,242,204}

\usepackage[a4paper,left=1.3cm, right=1.3cm, top=1.9cm, bottom=2.7cm,total={210mm,297mm}]{geometry}
\def\BibTeX{{\rm B\kern-.05em{\sc i\kern-.025em b}\kern-.08em
    T\kern-.1667em\lower.7ex\hbox{E}\kern-.125emX}}
    
\begin{document}

\title{From Buffers to Registers: Unlocking Fine-Grained FlashAttention with Hybrid-Bonded 3D NPU Co-Design}



\author{
  Jinxin Yu$^{1,2,*}$,
  Yudong Pan$^{1,2,*}$,
  \thanks{* indicates equal contribution to this work.}
  Mengdi Wang$^{1,2}$,
  Huawei Li$^{1,2}$,
  Yinhe Han$^{1,2}$,
  Xiaowei Li$^{1,2}$,
  Ying Wang$^{1,2}$\\[4pt]
  $^1$\textit{SKLP, Institute of Computing Technology, Chinese Academy of Sciences, Beijing, China}\\
  $^2$\textit{University of Chinese Academy of Sciences, Beijing, China}\\
  {\{yujinxin21s, panyudong23s, Wangmengdi, lihuawei, yinhes, lxw, wangying2009\}@ict.ac.cn}
}

\maketitle

\begin{abstract}

Transformer-based models dominate modern AI workloads but exacerbate memory bottlenecks due to their quadratic attention complexity and ever-growing model sizes. Existing accelerators, such as Groq and Cerebras, mitigate off-chip traffic with large on-chip caches, while algorithmic innovations such as FlashAttention fuse operators to avoid materializing large attention matrices. However, as off-chip traffic decreases, our measurements show that on-chip SRAM accesses account for over 60\% of energy in long-sequence workloads, making cache access the new bottleneck.
We propose 3D-Flow, a hybrid-bonded, 3D-stacked spatial accelerator that enables register-to-register communication across vertically partitioned PE tiers. Unlike 2D multi-array architectures limited by NoC-based router-to-router transfers, 3D-Flow leverages sub-10 µm vertical TSVs to sustain cycle-level operator pipelining with minimal overhead. On top of this architecture, we design 3D-FlashAttention, a fine-grained scheduling method that balances latency across tiers, forming a bubble-free vertical dataflow without on-chip SRAM round-trips.
Evaluations on Transformer workloads (OPT and QWEN models) show that our 3D spatial accelerator reduces 46–93\% energy consumption and achieves 1.4×–7.6× speedups compared to state-of-the-art 2D and 3D designs.

\end{abstract}


\input{Text/1.Introduction}

\input{Text/2.Background_and_motivation}

\input{Text/3.Architecture}

\input{Text/4.Dataflow}

\input{Text/5.Evaluation}

\input{Text/6.Conclusion}

\section*{Acknowledgment}

This work was supported by the National Natural Science Foundation of China (NSFC) under grant No.92373206. The corresponding authors are Ying Wang, Huawei Li and Xiaowei Li.


\bibliographystyle{IEEEtran} 
\newpage
\bibliography{ref} 
\end{document}

%% file: Text/1.Introduction.tex
\section{Introduction}

Large-scale AI inference, particularly Transformer-based large language models (LLMs)\cite{radford2019language,brown2020language,touvron2023llama,liu2024deepseek}, has become increasingly memory-intensive, where the cost of data movement dominates computation. Conventional accelerator~\cite{chen2016eyeriss,du2015shidiannao,liao2021ascend,gao2019tangram} designs devote significant effort to off-chip memory optimization through compression, prefetching, and bandwidth scaling, as well as to \textbf{operator fusion} techniques that reduce intermediate memory traffic. Among these, \textbf{FlashAttention}~\cite{dao2022flashattention,dao2023flashattention2fasterattentionbetter,shah2024flashattention3fastaccurateattention} represents a breakthrough: by restructuring attention into fused, tile-based operators, it eliminates the need to materialize the large $N\times N$ attention matrix and reduces off-chip DRAM traffic by more than an order of magnitude. When combined with large on-chip SRAM buffers, such fusion significantly alleviates the off-chip bandwidth bottleneck.

\begin{figure}
\centering
\includegraphics [width=0.7\linewidth]{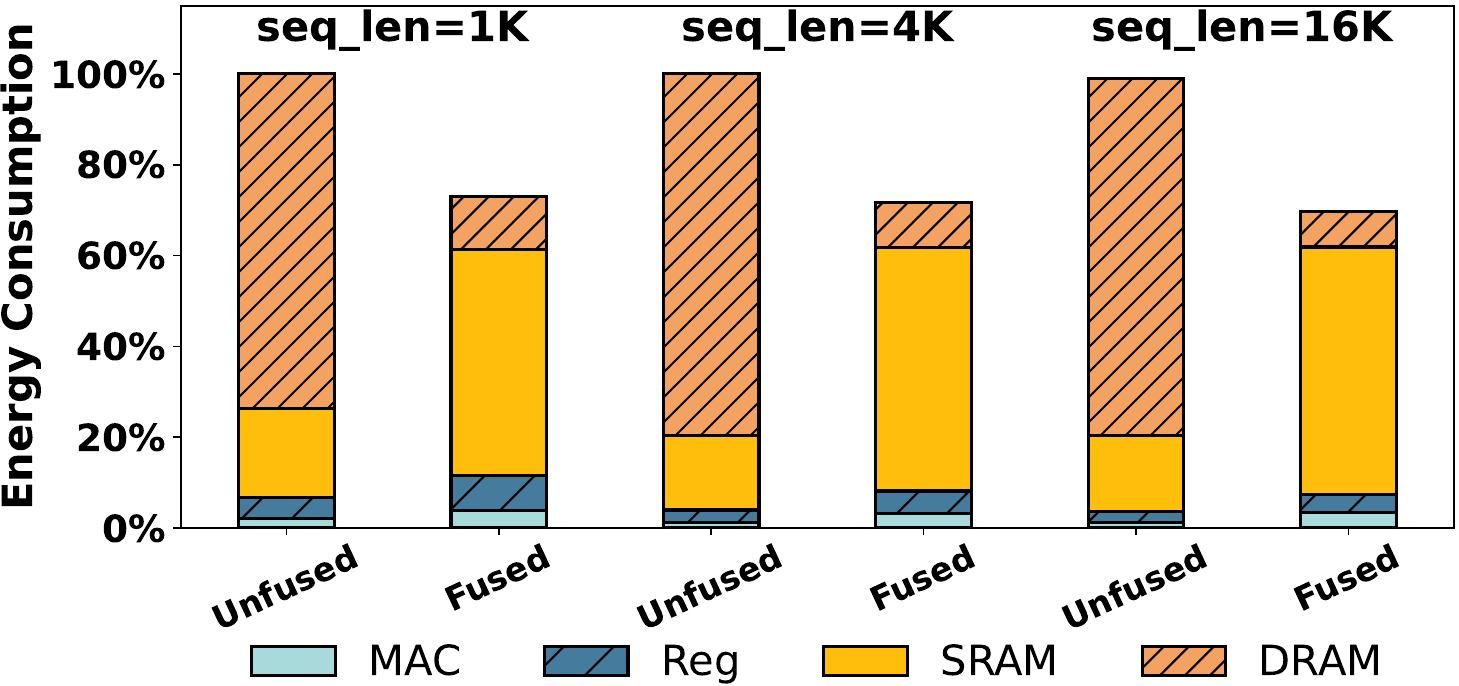}
\caption{Energy breakdown of operator fusion and unfusion with different sequence lengths for OPT.}
\label{fig:motivation}
\vspace{-0.3cm}
\end{figure}

However, as shown in Figure~\ref{fig:motivation}, our experiments reveal a new challenge: as off-chip memory traffic decreases, \textbf{on-chip SRAM accesses emerge as the dominant energy bottleneck}. Prior studies show that an SRAM access consumes $10–20\times$ more energy than a floating-point multiply-add (FMA)~\cite{horowitz20141}, and in FlashAttention workloads with long sequences ($N>2$k), on-chip buffer accesses account for over 60\% of total energy. This shift highlights that merely scaling on-chip capacity cannot sustain energy efficiency for future LLM accelerators.

One promising direction is to reduce on-chip storage reliance altogether, by spatially unrolling logical resources so that multiple fused operators can execute as a pipeline across processing elements (PEs). Inspired by FlashAttention’s fusion principle, we envision a \textbf{register-level pipeline} where operators such as $Q\times K^{T}$, $softmax$, and $S\times V$ are mapped onto different PEs and directly pass intermediate data without SRAM round-trips. This approach eliminates most on-chip buffer traffic and enables true dataflow-style execution.

Yet, realizing such fine-grained operator pipelining on large PE arrays faces two major obstacles: 1)\textbf{Inter-array connectivity}. Original FlashAttention fuses operators through \textbf{buffer-based data exchange} in on-chip SRAM, which still incurs high access cost. To minimize on-chip data movement, intermediate results must instead flow \textbf{register-to-register} across PEs, enabling cycle-accurate dataflow execution. However, existing 2D multi-array accelerators rely on \textbf{router-to-router NoC transfers} for inter-array communication. This indirection adds multi-cycle latency and energy overhead per hop, making it impossible to sustain fine-grained, cycle-level pipelining. The resulting \textbf{pipeline bubbles} cause resource underutilization, undermining the efficiency gains of operator fusion. 2) \textbf{Scheduling granularity}. The original FlashAttention fuses operators through \textbf{coarse-grained buffering}, but does not support register-to-register pipelining. Achieving \textbf{bubble-free}, \textbf{cycle-level} execution across operators requires a new fine-grained scheduling strategy.

To address these challenges, we propose \textbf{3D-Flow}, a three-dimensional systolic array architecture enabled by \textbf{hybrid bonding}~\cite{kagawa2019scaling,lu2024hybrid}. Unlike micro-bump interconnects with 40–50 µm pitch, hybrid bonding provides \textbf{sub-10 µm ultra-dense TSV} links with one-cycle latency and low energy per bit, enabling vertical PE-to-PE connections at scale. By vertically stacking PE tiers and assigning consecutive operators to adjacent tiers, 3D-Flow achieves cycle-level operator pipelining without requiring large SRAM buffers.

On top of this hardware substrate, we design \textbf{3D-FlashAttention}, a fine-grained scheduling strategy that partitions FlashAttention operators across tiers with \textbf{latency-balanced mapping}. This co-design ensures that each tier sustains comparable throughput, forming a tightly coupled vertical pipeline where intermediate results flow directly upward through TSVs. The result is a \textbf{bubble-free dataflow execution} that reduces both off-chip and on-chip traffic, significantly improving energy efficiency for Transformer inference.
In summary, this paper makes the following contributions:
\begin{itemize}
\item \textbf{Bottleneck analysis}. We demonstrate that FlashAttention on conventional 2D accelerators is limited by on-chip SRAM accesses, which consume over 60\% of energy in long-sequence workloads, and by operator load imbalance that reduces PE utilization.
\item \textbf{3D-Flow architecture}. We propose a hybrid-bonded, 3D-stacked systolic array where vertically partitioned PE tiers are connected with dense TSVs, supporting register-level operator pipelining with sub-cycle latency.
\item \textbf{3D-FlashAttention scheduling}. We introduce a fine-grained scheduling method that aligns operator latency across tiers, enabling bubble-free vertical pipelines and eliminating costly SRAM round-trips, thereby sustaining high throughput and energy efficiency for LLM inference. The co-designed Hybrid-bonded NPU architecture can also generalizable to other fused operators beyond attention.
\end{itemize}

%% file: Text/2.Background_and_motivation.tex
\section{Background and Motivation}
\subsection{FlashAttention: From GPU to Systolic Array}
FlashAttention~\cite{dao2022flashattention, dao2023flashattention2fasterattentionbetter, shah2024flashattention3fastaccurateattention} was developed as an IO-aware algorithm that minimizes data movement between high-latency off-chip DRAM and fast on-chip SRAM. It achieves this through two key techniques: tiling and operator fusion, as shown in algorithm~\ref{alg:fa2}. By partitioning the query, key, and value matrices into blocks that fit in SRAM, and fusing the $QK^{T}$, $softmax$, and $PV$ steps into a single kernel, FlashAttention avoids writing large intermediate tensors back to DRAM and achieving exceptional performance on NVIDIA A100 and H100 GPUs. However, translating these gains to accelerators built on systolic arrays presents significant challenges that offer considerable optimization potential. 

A major bottleneck arises from load imbalance between heterogeneous compute units: $QK^{T}$ multiplication fully utilizes high-throughput systolic arrays, whereas the subsequent $softmax$ runs on slower scalar or vector units, causing stalls and underutilization of matrix hardware. In addition, on-chip memory bandwidth is constrained, as data transfer between large caches and systolic arrays is serialized over multiple cycles, unlike the parallel shared-memory access of GPUs. This high-latency transfer scales with cache size, making memory access a dominant cost. Overcoming these limitations requires a co-designed architecture and dataflow.


\begin{algorithm}[htbp]
\small 
\linespread{.9}\selectfont 
\caption{FlashAttention-2 forward pass}
\label{alg:fa2}
\KwIn{Matrices $Q,K,V \in \mathbb{R}^{\text{LEN} \times d}$}

Divide $Q$ into $T_r = \lceil \frac{\text{LEN}}{B_r} \rceil$ blocks of size $B_r\times d$ each, and divide $K$ and $V$ into $T_c = \lceil \frac{\text{LEN}}{B_c} \rceil$ blocks of size $B_c\times d$ each\;

\For{$1 \le i \le T_r$}{
  Initialize $old_m, old_l = (-\infty), (0) \in \mathbb{R}^{B_r}$\;
  Initialize $old_O = (0) \in \mathbb{R}^{B_r\times d}$\;
  
  \For{$1 \le j \le T_c$}{
    \setlength{\fboxsep}{0pt}\colorbox{myred}{$S \gets Q_iK_j^\top \in \mathbb{R}^{B_r\times B_c}$}\;
    \colorbox{mygreen}{$local_m \gets \mathrm{rowmax}(S) \in \mathbb{R}^{B_r}$}\;
    \colorbox{mygreen}{$new_m \gets \max(local_m, old_m) \in \mathbb{R}^{B_r}$}\;
    \colorbox{mygreen}{$a \gets old_m - new_m \in \mathbb{R}^{B_r}$}\;
    \colorbox{myblue}{$b \gets \exp(\frac{a}{\sqrt{d}}) = \exp2(\frac{\log_2 e}{\sqrt{d}} a) \in \mathbb{R}^{B_r}$}\;
    \colorbox{mygreen}{$N \gets S - new_m \in \mathbb{R}^{B_r\times B_c}$}\;
    \colorbox{myblue}{$P \gets \exp(\frac{N}{\sqrt{d}}) = \exp2(\frac{\log_2 e}{\sqrt{d}} N) \in \mathbb{R}^{B_r\times B_c}$}\;
    \colorbox{myblue}{$local_l \gets \mathrm{rowsum}(P) \in \mathbb{R}^{B_r}$}\;
    \colorbox{myblue}{$new_l \gets old_l \times b + local_l \in \mathbb{R}^{B_r}$}\;
    
   \colorbox{myyellow}{$local_O \gets PV_j \in \mathbb{R}^{B_r\times d}$}\;
   \colorbox{myyellow}{$new_O \gets \mathrm{diag}(b)old_O + local_O \in \mathbb{R}^{B_r\times d}$}\;

    $old_m \gets new_m$\;
    $old_l \gets new_l$\;
    $old_O \gets new_O$\;
  }
  $O_i \gets \mathrm{diag}(old_l)^{-1} old_O \in \mathbb{R}^{B_r\times d}$\;
}
\end{algorithm}

Several prior works have carried out early-stage investigations into 2D PE arrays. COSA\cite{wang2023cosa} maps FlashAttention onto two collaborating systolic arrays, yet the softmax computation still relies on a dedicated special function unit (SFU). This unit may not deliver enough FLOPs/s to match the throughput of the systolic arrays. FuseMax\cite{nayak2024fusemax} was the first to integrate FlashAttention entirely within a single systolic array. It implements the rowmax and rowsum steps as spatial reduction operations and concurrently pipelines multiple FlashAttention iterations on the same array. Nevertheless, to enable context switching between iterations, each PE must store intermediate values in 10 registers, resulting in extra register usage.
Despite these efforts, the fundamental limitations of 2D PE arrays remain unresolved. In FlashAttention, consecutive operators still rely on the on-chip SRAM for intermediate result exchange, incurring significant and unnecessary memory access overhead. Furthermore, load imbalance among consecutive operators continues to cause underutilization of computational resources and processing stalls.

\subsection{Opportunities Enabled by 3D Integration}

Recent advances in semiconductor packaging, exemplified by hybrid bonding, offer a promising path to address the intrinsic dataflow limitations of conventional 2D integration. Traditional planar interconnects suffer from limited wiring density, high latency, and excessive power when scaling to data-intensive workloads. In contrast, 3D integration enables vertical stacking of multiple dies, shortening interconnect distances and increasing inter-die bandwidth density. Hybrid bonding~\cite{kagawa2019scaling,lu2024hybrid} directly connects copper pads at the wafer or die level, achieving ultra-fine-pitch ($<10\,\mu\text{m}$) vertical links with lower resistance and capacitance than TSV- or micro-bump-based approaches~\cite{wang2023short}. These low-parasitic interconnects reduce transmission delay, raise signaling rates, and improve energy efficiency. The benefits of this packaging paradigm have already been realized in commercial products ,including high-bandwidth memory (HBM) stacks\cite{jun2017hbm} and 3D SRAM caches such as AMD's 3D V-Cache\cite{wuu20223d}.

Beyond memory-on-logic stacking, 3D integration allows vertically partitioning the compute fabric itself. Mapping consecutive operators, such as attention sub-stages, onto densely interconnected PE tiers via hybrid-bonded links enables intermediate results to be transferred directly between layers without traversing on-chip SRAM. This vertically stacked architecture reduces on-chip SRAM traffic, and supports operator-specific resource allocation across tiers, thereby alleviating operator load imbalance and sustaining high throughput and energy efficiency for memory-intensive workloads.

\subsection{Motivation}
Overall, there are two factors that lead to the inefficiency of running FlashAttention on a 2D systolic array. \textbf{First}, due to the physical constraints of the PE array, intermediate results between consecutive operators must be exchanged through the on-chip cache, which creates a bottleneck in on-chip memory accesses. \textbf{Second}, load imbalance between consecutive operators frequently leads to idle cycles and underutilization of compute resources. These two factors hinder the efficient execution of FlashAttention on existing 2D PE array based accelerators, which in turn motivates us to explore emerging 3D integration technologies. Through a co-design of architecture and dataflow, we can vertically stack PE arrays with different functionalities, and carefully orchestrate the workload so that consecutive operators are evenly grouped according to load balance and mapped onto the 3D array. In this way, intermediate results can be directly transferred to the next operator via fine-pitch TSV-link enabled by hybrid bonding, significantly eliminating the on-chip memory bottleneck and improving resource utilization, and thereby unlocks the full performance potential of FlashAttention in hardware accelerators.

%% file: Text/3.Architecture.tex
\section{3D-Flow Architecture}

\begin{figure}[tb]
\centering
\includegraphics [width=1\linewidth]{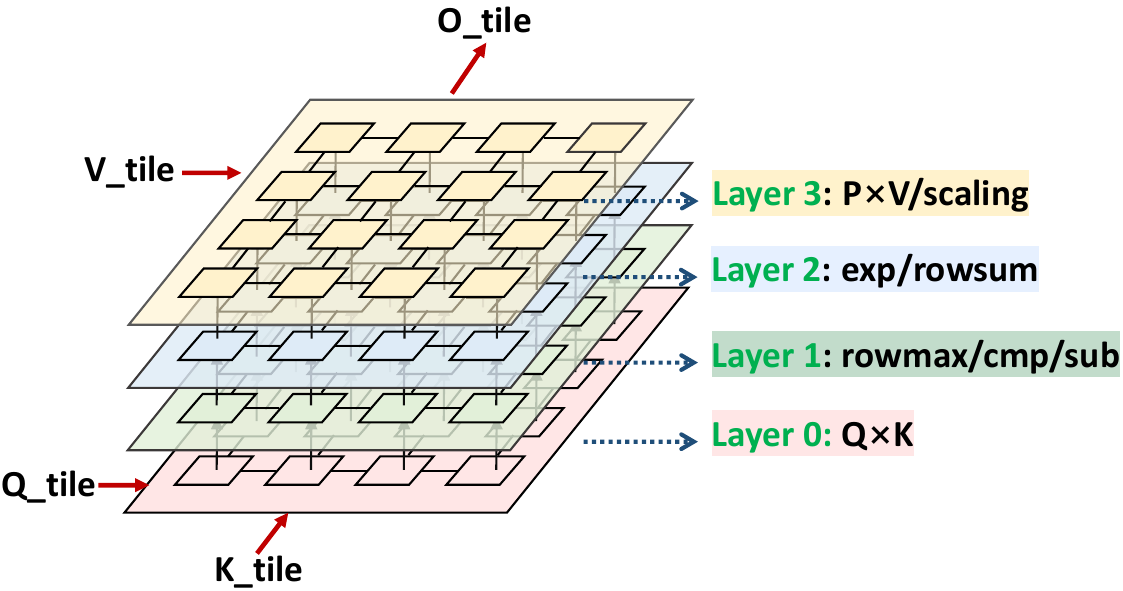}
\caption{Overview of 3D-stacked PE array architecture and the operator mapping of each layer.}
\label{fig:architecture}
\vspace{-1em}
\end{figure}

\begin{figure}[htbp]
    \centering
    \vspace{-0.2cm}
    \begin{minipage}{0.23\textwidth}
        \centering
        \includegraphics[width=\linewidth]{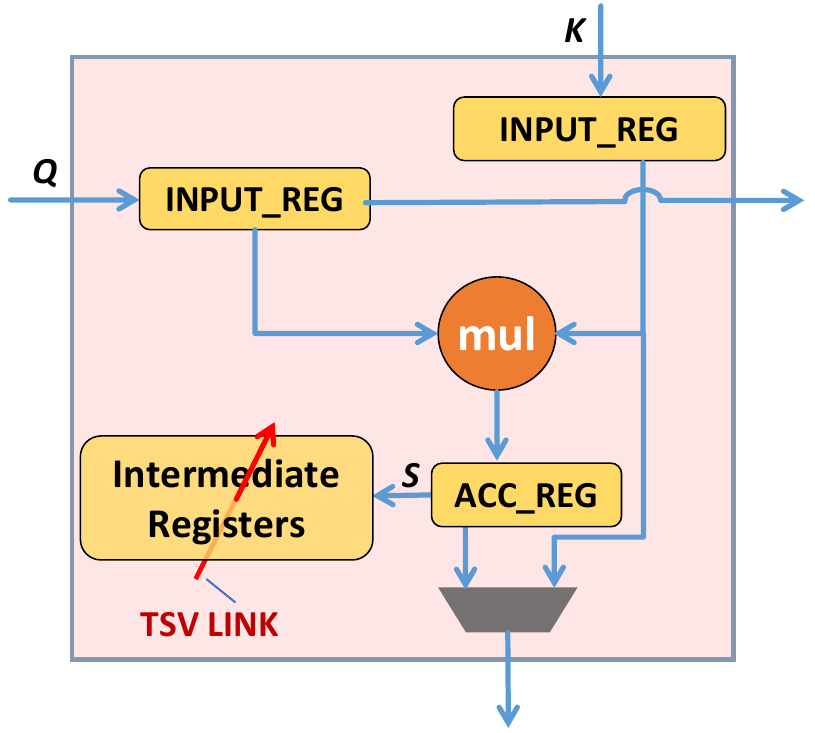}
        \subcaption{PE in layer\_0.}
        \label{fig:pe0}
    \end{minipage}
    \hfill
    \begin{minipage}{0.23\textwidth}
        \centering
        \includegraphics[width=\linewidth]{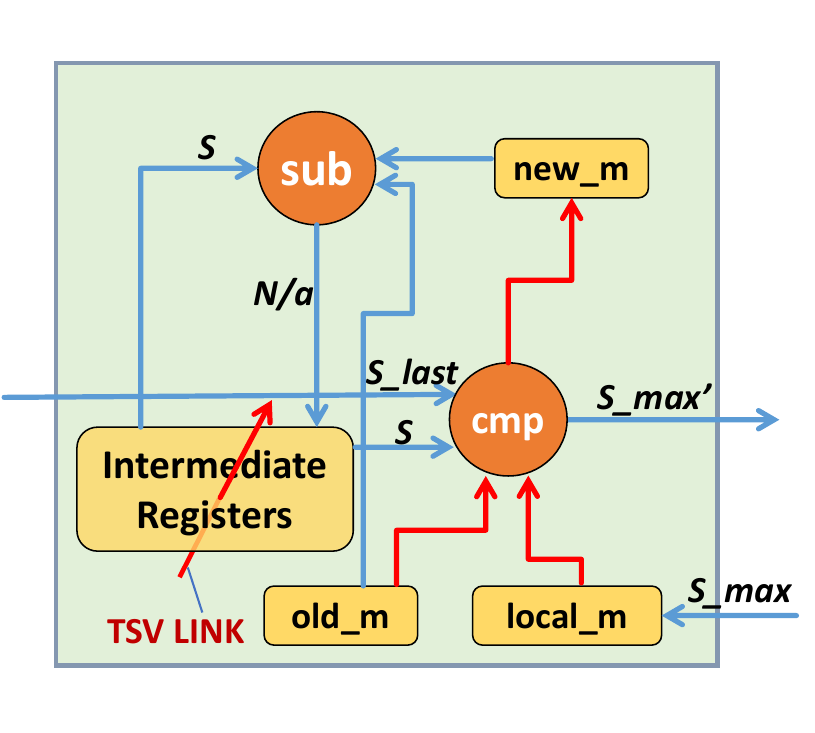}
        \subcaption{PE in layer\_1.}
        \label{fig:pe1}
    \end{minipage}
    
    \vspace{0.1cm} 
    
    \begin{minipage}{0.23\textwidth}
        \centering
        \includegraphics[width=\linewidth]{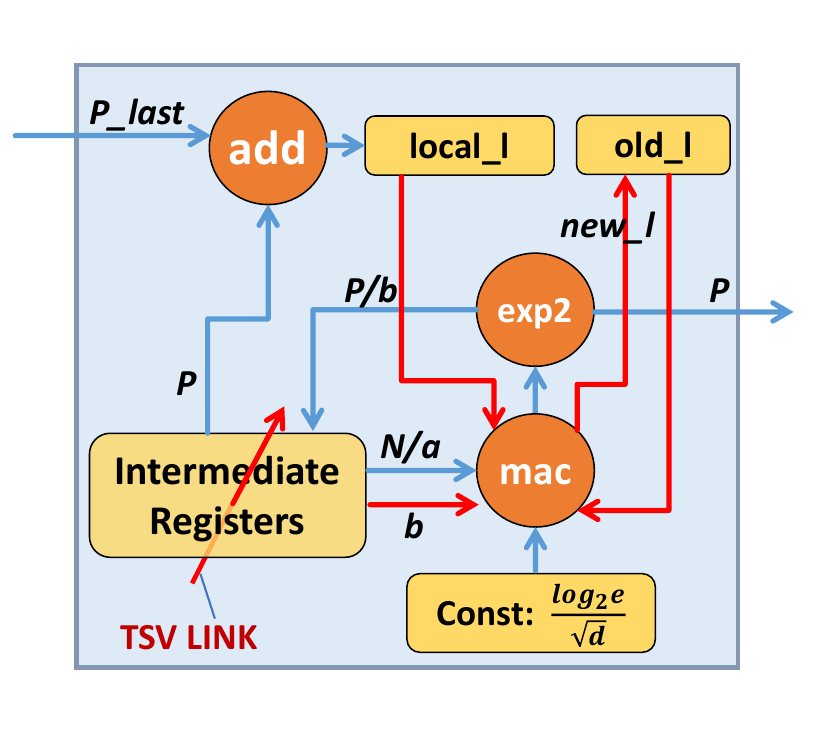}
        \subcaption{PE in layer\_2.}
        \label{fig:pe2}
    \end{minipage}
    \hfill
    \begin{minipage}{0.23\textwidth}
        \centering
        \includegraphics[width=\linewidth]{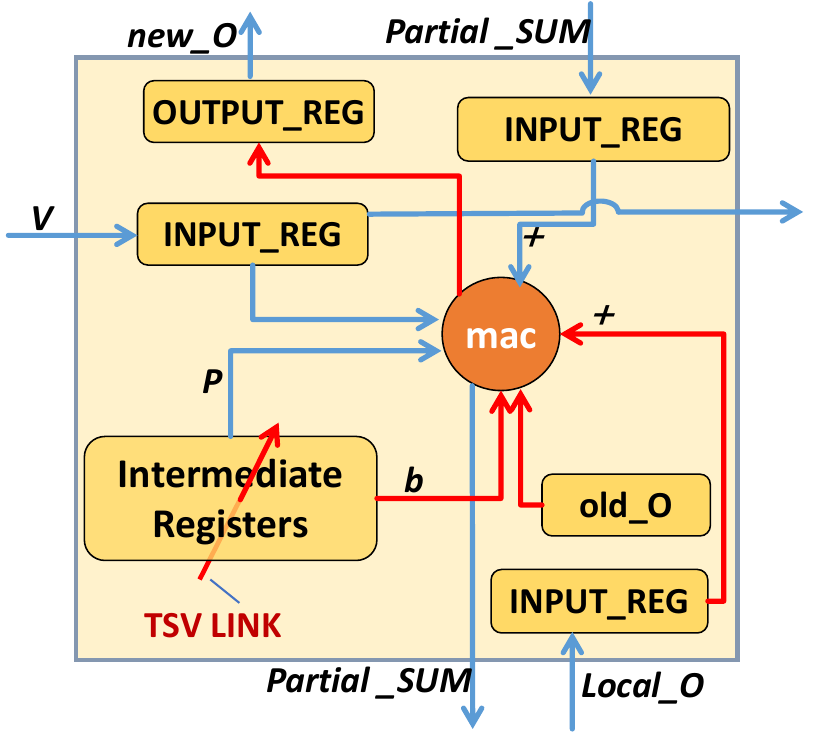}
        \subcaption{PE in layer\_3.}
        \label{fig:pe3}
    \end{minipage}
    \caption{Custom Design of PE Units in Each Layer.}
    \label{PE Designs}
    \vspace{-0.2cm}
\end{figure}

\begin{figure*}[htbp]
    \centering
    \vspace{-0.2cm}
    \begin{minipage}{0.5\textwidth}
        \centering
        \includegraphics[width=\linewidth]{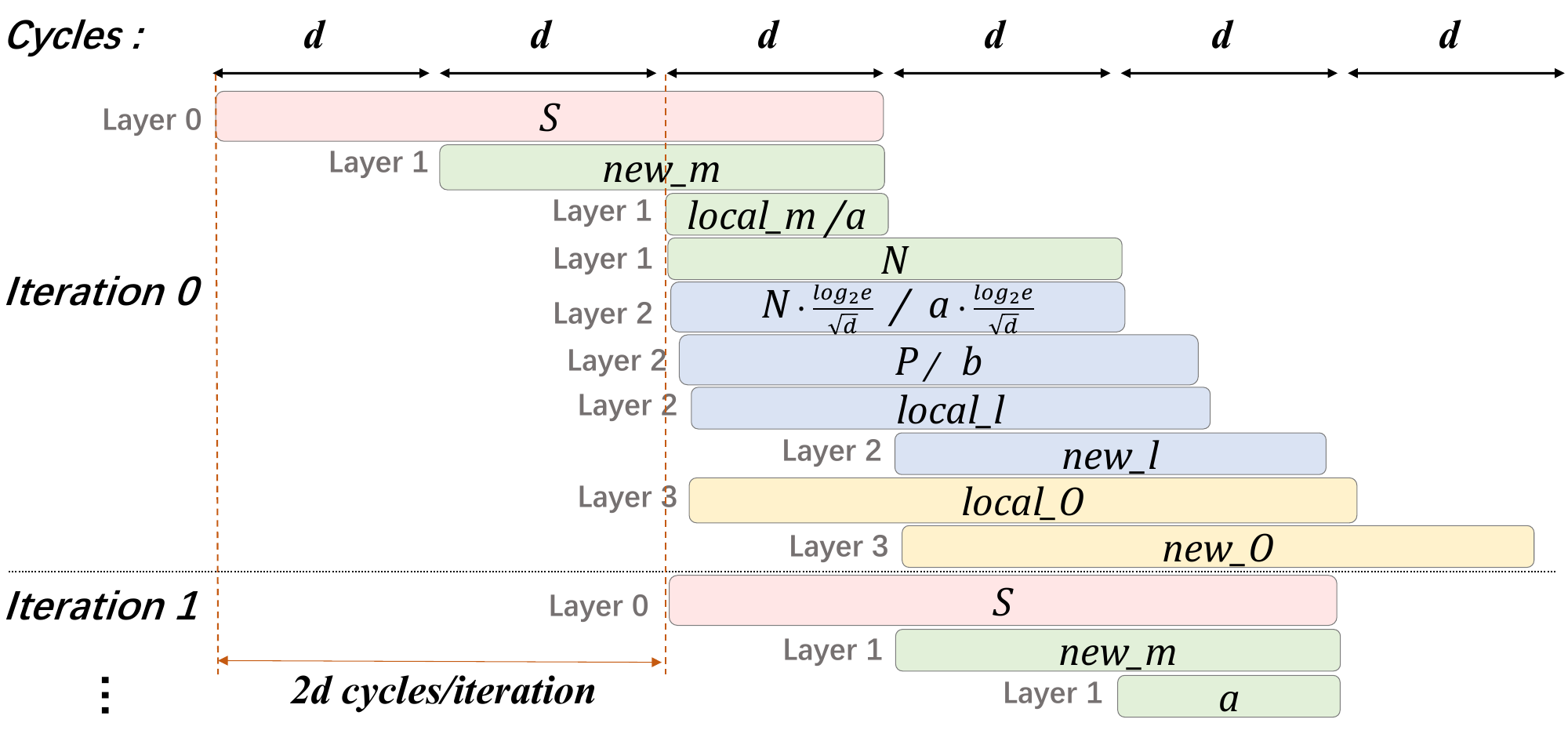}
        \subcaption{Operator pipeline.}
        \label{fig:operator flow}
    \end{minipage}
    \hfill
    \begin{minipage}{0.23\textwidth}
        \centering
        \includegraphics[width=\linewidth]{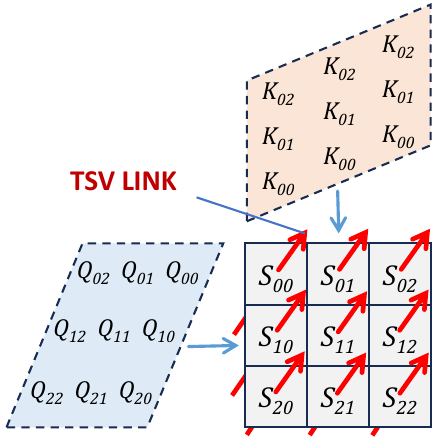}
        \subcaption{$QK^{T}$ dataflow.}
        \label{fig:QK}
    \end{minipage}
    \hfill
    \begin{minipage}{0.23\textwidth}
        \centering
        \includegraphics[width=\linewidth]{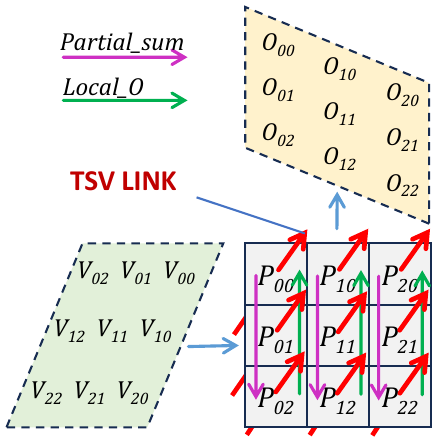}
        \subcaption{$PV$ dataflow.}
        \label{fig:PV}
    \end{minipage}
    \caption{(a) 3D dataflow and pipeline of consecutive FlashAttention operators across 3D PE array. Operator blocks with different colors correspond to the rows with the same background in Algorithm~\ref{alg:fa2}. Each iteration corresponds to one inner loop execution of FlashAttention. $d$ denotes the number of rows/columns in each PE array. When the pipeline is fully filled, one iteration can be completed in $2d$ cycles. (b) Input dataflow of 3D array. (c) Output dataflow of 3D array.}
    \label{dataflow}
    \vspace{-0.2cm}
\end{figure*}

\subsection{Overview}
Figure \ref{fig:architecture} illustrates the overview of the proposed 3D-stacked PE array architecture. Adjacent compute tiers are integrated in a face-to-back configuration, where the registers of spatially aligned PEs are directly interconnected via hybrid-bonded TSV links to enable low-latency data forwarding.  Layers with different colors are responsible for distinct operators in FlashAttention, corresponding to the code lines with the same background colors in Algorithm~\ref{alg:fa2}. The pipeline of operator flow is carefully orchestrated to ensure balanced workloads across layers. The PE units structure in each layer are custom-designed according to the functionality of the corresponding operators, as shown in Figure~\ref{PE Designs}. A bottom-to-top computation flow corresponds to one iteration of the inner loop in the Algorithm~\ref{alg:fa2}, enabling multiple iterations to be executed in parallel through the systolic movement of data within the 3D array. Each PE array layer is configured with dimensions of $d \times d$, where $d$ represents the attention head dimension, thereby allowing a each layer to complete the computation of a whole tile.


\subsection{PE Design}
Figure~\ref{PE Designs} illustrates the detailed microarchitecture of the PE units, each custom-designed for the operators executed in its respective layer. The intermediate registers of corresponding PEs in adjacent tiers are interconnected via TSV links, enabling single-cycle forwarding of intermediate results and avoiding the high-latency SRAM routing in conventional 2D designs. 
As illustrated in Figure~\ref{PE Designs}, PE in Layer~0 integrates MAC units and is configured with an Outer-Stationary (OS) dataflow to execute the $QK^T$ computation. 
PE in Layer~1 integrates comparator modules and adders (configured for subtraction) to determine row-wise maximum values and perform subtraction operations, supporting operators such as \textit{RowMax}. 
PE in Layer~2 integrates exponential computation units, multipliers, and adders, primarily responsible for operators including \textit{exp} and \textit{RowSum}. 
PE in Layer~3 integrates MAC units along with a reverse data path to implement the $PV$ operation and the scaling of the output results. 
The detailed execution procedures of each operator within the PEs, along with the corresponding dataflow, will be detailed in Section~\ref{sec:Layer-wise Dataflow}.


\subsection{Thermal Feasibility}
We estimated the thermal behavior of stacked $128 \times 128$ PE array. Based on synthesis results, the single PE array has an estimated area of about $A \approx 80~\text{mm}^2$. At peak activity ($P_{PE}=200~\mu\text{W}$), the power of a PE array is $P_{\text{layer}} \approx 3.3~\text{W}$, and the four-layer stack consumes $P_{\text{total}} \approx 13.1~\text{W}$, yielding a layer power density of $\rho \approx 41~\text{W/cm}^2$. Using a first-order thermal resistance model, the internal temperature rise is small ($\sim$2.8°C), indicating good vertical heat conduction. Under a conservative assumption of package thermal resistance $R_{\theta JA} \approx 2.5~\text{K/W}$~\cite{Semiconductor}, the junction temperature at 25°C ambient is approximately $T_j \approx 83^\circ\text{C}$. With standard advanced packaging techniques, such as thermal TSVs~\cite{wang2024review}, $R_{\theta JA}$ can be further reduced, keeping peak temperatures well within safe operating limits.

%% file: Text/4.Dataflow.tex
\section{3D-Flashattention Dataflow}
\subsection{Overview}

As shown in Figure~\ref{fig:operator flow}, the consecutive FlashAttention operators are pipelined across different layers of the 3D-stacked PE array. Each block represents a distinct operator, and blocks with the same color are executed in the same layer. The colors of the operators correspond to the background colors of different rows in Algorithm~\ref{alg:fa2}. Each iteration corresponds to one inner loop in FlashAttention: given an input $d \times d$ $QKV\_tile$, a $d \times d$ attention result is produced, where $d$ denotes both the row/column size of the PE array, which is also equal to the dimension of an attention head. Through careful scheduling, operators are mapped to different layers based on their functionality and computational load, and co-designed with the PE microarchitecture to minimize inter-operator bubbles. When the pipeline is fully filled, only $2d$ cycles are required to complete an iteration, whereas a conventional 2D PE array needs $3d$ cycles merely to perform the $QK^{T}$ operation. The following section details the execution of the operator flow on the 3D PE array.

\subsection{Layer-wise Dataflow}\label{sec:Layer-wise Dataflow}
\subsubsection{\textbf{Layer\,0: $\mathbf{QK^{T}}$}}
As shown in Figure~\ref{fig:QK}, $Layer\_0$ executes the $QK^{T}$ operator by feeding $Q\_tile$ and $K\_tile$ into the systolic array from horizontal and vertical directions in a parallelogram injection pattern for MAC operations. Partial sums are accumulated within PEs, and after $d$ cycles, $S_{\mathrm{(0,0)}}$ is completed and forwarded via TSV to the corresponding PE in $Layer\_1$. Thereafter, new $S$ element is produced per cycle, with all results ready after $3d$ cycles. Meanwhile, as illustrated in Figure~\ref{fig:operator flow}, the next iteration of the $QK^{T}$ operator can start after 2d cycles, once the top-left PE becomes idle.

\subsubsection{\textbf{Layer\,1: Row-wise Maximum and Subtraction}}
The primary function of this layer is to compute the row-wise maximum $S_{\mathrm{max}}$ of each row in the matrix $S$ and subtract it. As shown in Figure~\ref{fig:operator flow}, after $d$ cycles, the top-left PE receives the first $S_{(0,0)}$ via the TSV\_link. Thereafter, in each cycle it propagates rightward, as shown in Figure~\ref{fig:pe1}, where it is compared with the newly received $S$ in the current PE to obtain a larger value $S_{\mathrm{max}}'$, and then continued to be propagated rightward. At $2d$ cycles, the $S_{\mathrm{max}}$ of the first row is obtained. Then $S_{\mathrm{max}}$ is propagated leftward in the first row, compared with $old\_m$, and the larger value is updated to $new\_m$. Subtraction with both $S$ and $old\_m$ yields $N$ and $a$, respectively. Once $N$ and $a$ are computed, they are transmitted via the TSV\_link to the corresponding PE in Layer\,2 for the execution of the next operator. Since $a$ exists only in the rightmost PE column, the vector $a$ and the matrix $N$ are completed at $3d$ and $4d$ cycles, respectively.

\subsubsection{\textbf{Layer\,2: Exponential-related Operations}}
This layer is mainly responsible for the element-wise exponential computations in the \textit{softmax} operation, as well as multiply-accumulate operations involving the scale variable. As shown in Figure~\ref{fig:operator flow}, at $2d$ cycles, the first value of the matrix $N$ has been computed and is immediately transmitted via the TSV\_link to this layer for subsequent processing. For the exponential function, we employ the more efficient \texttt{exp2} implementation commonly used in hardware accelerators~\cite{hussain2022area}. As illustrated in Figure~\ref{fig:pe2}, the constant $\frac{\log_{2} e}{\sqrt{d}}$ is stored in the register of each PE and first multiplied with $N/a$, and the result is then passed to the exponential unit to obtain $P/b$. Afterwards, $P$ is sent in one direction to the intermediate register and then forwarded via the TSV\_link to the corresponding PE in Layer\,3, and in the other direction it propagates leftward for accumulation with $P$ from the neighboring PE to produce $local\_l$. Finally, $local\_l$ undergoes a multiply-accumulate with $b$ and $old\_l$ to generate $new\_l$, after which $old\_l$ is updated. This layer contains multiple operators that can be executed in parallel, starting from the $2d$ cycle and completing before the $5d$ cycle.

\subsubsection{\textbf{Layer\,3: $\mathbf{PV}$ and $\mathbf{O}$ Scaling}}
This layer is primarily responsible for executing the $PV$ operator and scaling the current output $local\_O$ using the previous iteration's result $old\_O$ in the PE’s local registers. As shown in ~\ref{fig:operator flow}, at the $2d$ cycle, $V\_tile$ is fed into the PE array from the left following a parallelogram pattern, and is multiplied with the matrix $P$ received via the TSV\_link under a \textit{Weight-Stationary} (WS) dataflow. Partial sums are propagated downward along the vertical direction for accumulation. Once the last row produces $local\_O$, the data is transmitted upward to the first row and undergoes a multiply-accumulate operation with $old\_O$ and $d$ to generate $new\_O$. The final results are then output from the top of the array. In this layer, the $PV$ computation starts at the $2d$ cycle, and the first $local\_O$ is obtained at the $3d$ cycle, after which the $O$ scaling operation is initiated, and all computations are completed at the $5d$ cycle.

Through the aforementioned bubble-free vertical pipelines, a complete inner-loop operator flow of FlashAttention can be completed every $2d$ cycles. Intermediate results between operators are directly transferred via \textit{TSV\_links} between registers, thereby avoiding costly SRAM round-trips. The iteration counts of the inner and outer loops in Algorithm~\ref{alg:fa2} are determined by the ratio between the attention sequence length and $d$. Upon completion of the inner loop, an additional operation is performed, as shown in line~21 of Algorithm~\ref{alg:fa2}, to scale $O$ using $old\_l$. When the outer loop finishes, the computation of a single attention head is complete. Multiple heads can be processed in parallel by integrating multiple 3D-stacked PE arrays, depending on the target application requirements and hardware budget.

%% file: Text/5.Evaluation.tex
\section{Evaluation}
\subsection{Experiment Setup}

\textbf{Accelerator modeling methodology.} We evaluate performance and energy of our 3D architecture and baselines using a cycle-accurate in-house simulator, which integrates Ramulator~\cite{luo2023ramulator} to capture DRAM access latency and bandwidth, and a custom systolic array simulator to model the specific dataflow we designed for the four-tier architecture. A double-buffering mechanism between memory levels is included to hide access latency. Energy is estimated by tracking hardware activity and invoking Accelergy~\cite{wu2019accelergy}. To ensure design feasibility, we implemented the complete four-tier PE array in RTL and synthesized it with Synopsys Design Compiler~\cite{synopsys_dc} on TSMC 16nm, verifying the functional correctness of our 3D-FlashAttention dataflow and calibrating the energy model. For inter-tier register-to-register communication overhead, we conservatively set the Z-axis data transfer energy~\cite{rich2023thermal} to 1.35 pJ/byte, based on stacked DRAM analysis~\cite{mathur2021thermal}, providing a reasonable upper-bound for on-chip register transfers.

\textbf{Baselines.} To comprehensively evaluate the advantages of our proposed 3D architecture and dataflow, we compare it against a series of 2D and 3D baselines. \textbf{(1) 2D-Unfused}: A conventional systolic array that executes the phases of attention computation sequentially without operator fusion. \textbf{(2) 2D-Fused}: Representative designs, such as FuseMax~\cite{nayak2024fusemax}, FLAT~\cite{kao2023flat}, and TileFlow~\cite{zheng2023tileflow}, which perform deep operator fusion on a single array. \textbf{(3) Dual-SA}: A 2D dual-array architecture inspired by COSA~\cite{wang2023cosa}. It partitions the $QK^{T}$ and $PV$ computations onto two arrays connected by a dedicated Softmax unit, enabling intermediate results to flow directly between these compute units while bypassing on-chip SRAM. \textbf{(4) 3D-Base}: An accelerator architecturally identical to ours but maps each operator onto the stacked tiers following the methods in~\cite{joseph2021architecture,kung2018mapping}, with intermediate results exchanged via SRAM. To ensure a fair comparison, all accelerators are configured with equivalent compute and storage resources (detailed in Table \ref{tab:Accelerator Specification in Evaluation}) and are provisioned with sufficient DRAM. For baselines that require a mapping search (i.e., 2D-Unfused and 3D-Base), we developed a cost model based on Timeloop~\cite{parashar2019timeloop} and utilized linear pruned search strategy to find optimal tiling schemes.


\begin{table}[h]
    \centering
    \caption{Accelerator Specification in Evaluation}
    \renewcommand{\arraystretch}{1.5}
    \resizebox{\columnwidth}{!}{
    \begin{tabular}{|c|c|c|c|}
        \hline
         & \rule{0pt}{2.2em}\shortstack{Ours \\ 3D-Base} 
         & \rule{0pt}{2.2em}\shortstack{2D-Unfused \\ 2D-Fused} 
         & \rule{0pt}{2.2em}Dual-SA  \\ \hline
        \textbf{Array Size} & $128\times128\times4$ & $128\times128$ & $128\times128\times2$ \\ \hline
        \textbf{Clusters} & 1 & 4 & 2 \\ \hline
        \textbf{On-Chip Mem. Size} & 60MB & 60MB & 60MB \\ \hline
        \textbf{On-Chip BW} & 8 TB/Sec & 8 TB/Sec & 8 TB/Sec \\ \hline
        \textbf{Off-Chip BW} & 400GB/Sec & 400GB/Sec & 400GB/Sec \\ \hline
    \end{tabular}
    }
    \renewcommand{\arraystretch}{1}
    \label{tab:Accelerator Specification in Evaluation}
\end{table}


\textbf{Benchmarks.} We evaluate the accelerators using representative Transformer models to cover different attention mechanisms: OPT~\cite{zhang2022opt} for Multi-Head Attention (MHA) and Qwen~\cite{bai2023qwen} for Grouped-Query Attention (GQA). The evaluation spans sequence lengths from 1K to 64K, enabled by implementing dynamic RoPE scaling~\cite{shang2025longrope2} to extend the models' context windows beyond their pre-trained limits.


\subsection{Experimental Results}
\subsubsection{\textbf{Energy consumption}}

\begin{figure}[ht]
\centering
\includegraphics[width=1\linewidth]{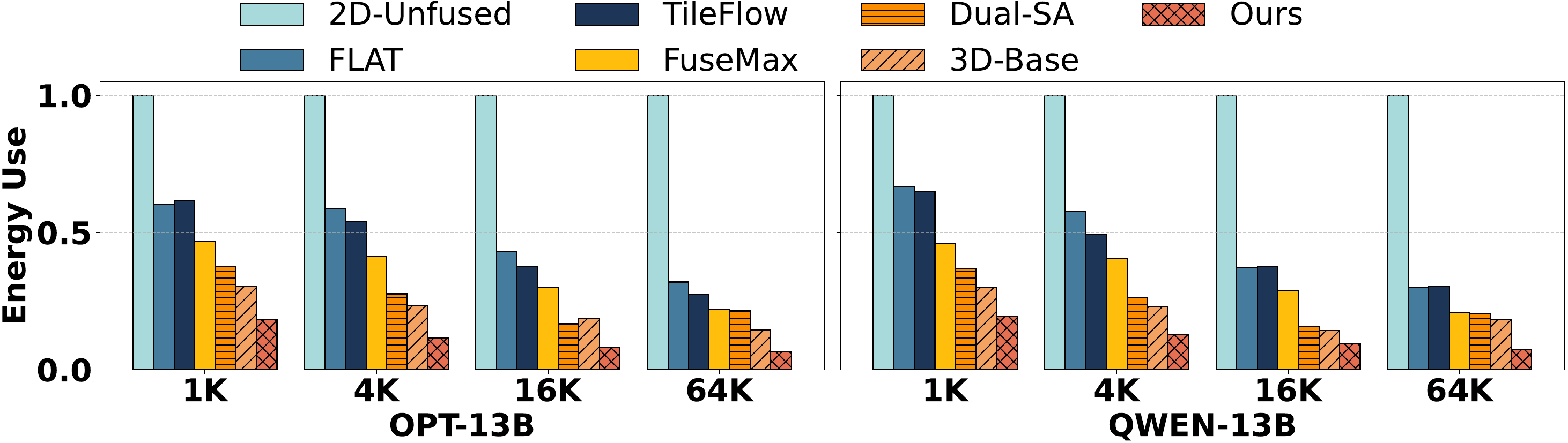}
\caption{Comparison of energy consumption of attention with sequence lengths ranging from \textit{1K} to \textit{64K}.}
\label{fig:energy_comparison}
\end{figure}

Figure~\ref{fig:energy_comparison} compares the energy consumption of our design against several baselines for the attention computation, with all results normalized to the 2D-Unfused baseline. Overall, our design demonstrates significant energy efficiency, reducing energy by 80.5\% to 93\% compared to 2D-Unfused approach. Against advanced 2D fusion schemes (FuseMax and Dual-SA), it saves 54.2\% to 66.7\% in energy. Even compared to the architecturally identical 3D-Naive baseline, our specialized dataflow provides an average energy advantage of 46.8\%. This advantage stems from our design's significant reduction in both costly off-chip DRAM and the increasingly critical on-chip SRAM accesses.

\begin{figure}[ht]
\centering
\includegraphics [width=1.0\linewidth]{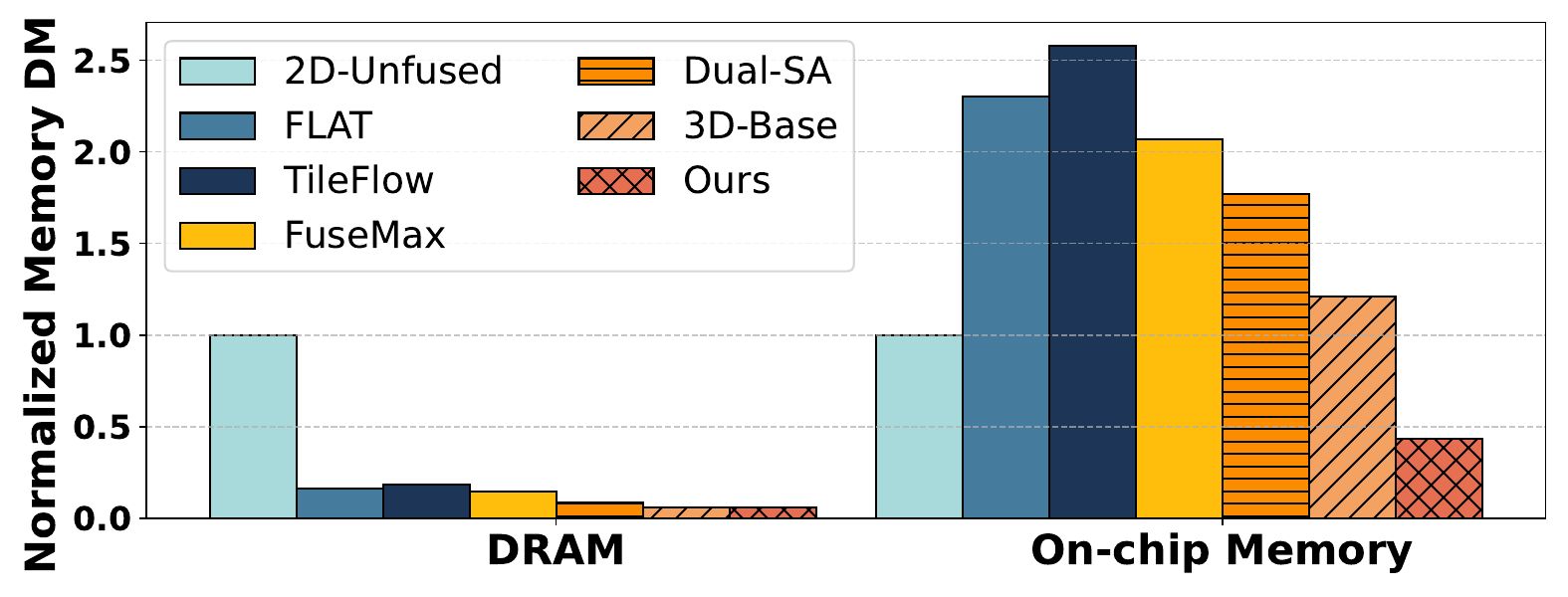}
\caption{Comparison of average memory DM(data movement volume) of attention.}
\label{fig:memory_dm}
\end{figure}

Figure~\ref{fig:memory_dm} breaks down data movement of each baseline and reveals the sources of energy savings. Compared to 2D-Unfused, operator fusion methods like FuseMax reduce DRAM accesses by avoiding the materialization of the large intermediate attention score matrix, quadratic in sequence length. However, this shifts the bottleneck to on-chip memory; while FuseMax cuts DRAM accesses by 85.5\%, it increases on-chip SRAM traffic by $2.1\times$, creating a new bottleneck. The Dual-SA architecture attempts to alleviate this with its dual-array design, but its dedicated Softmax unit still relies on SRAM for data exchange, preventing a seamless dataflow between PE arrays. Meanwhile, the 3D-Base baseline leverages 3D stacking for some input reuse (e.g., broadcasting Q-tiles across tiers) but, without a co-designed dataflow, still requires SRAM as an intermediary between operators, leaving 3D integration potential untapped.

In contrast, our design leverages 3D-stacked architecture with a co-designed, specialized 3D-FlashAttention dataflow that maps different computational stages to distinct vertical PE array tiers. This strategy not only reduces DRAM accesses via fusion but, more critically, enables intermediate results to flow directly between tiers via registers, completely bypassing on-chip SRAM. Our design reduces on-chip SRAM traffic by an average of 76.6\% compared to the other fusion baselines and realizing a truly compute-centric dataflow. Since attention dominates end-to-end energy in long sequences~\cite{dao2022flashattention}, our design yields substantial inference savings, reducing overall energy by 32.7\% to 64.2\% on average compared to baselines.


\begin{table}[h!]
\centering
\caption{Average Energy Breakdown of Our Design}
\label{tab:energy_breakdown}
\resizebox{\columnwidth}{!}{ 
\begin{tabular}{@{}lccccc@{}}
\toprule
\textbf{Component} & \textbf{MAC} & \textbf{Reg} & \textbf{SRAM} & \textbf{DRAM} & \textbf{3D-IC Overhead} \\ \midrule
seq\_len=\textit{1K}        & 8.5\%      & 21.2\%        & 38.3\%       & 26.7\%       & 5.3\%           \\ 
seq\_len=\textit{4K}        & 11.7\%     & 31.9\%        & 35.0\%       & 15.1\%        & 6.3\%           \\
seq\_len=\textit{16K}       & 10.4\%     & 29.2\%        & 29.5\%       & 20.8\%       & 10.1\%           \\ 
seq\_len=\textit{64K}       & 12.0\%     & 34.4\%        & 28.5\%       & 16.2\%       & 8.9\%             \\ \bottomrule
\end{tabular}
}
\end{table}

\subsubsection{\textbf{Energy breakdown and overhead}} 
We conduct detailed modeling of the energy breakdown for our 3D architecture across various sequence lengths. As shown in Table \ref{tab:energy_breakdown}, the energy consumption is primarily driven by memory access at different levels and is relatively evenly distributed. Our design incurs additional overhead from increased register access and an average energy overhead of 7.81\% due to the use of 3D-IC technology. However, these overheads remain well within acceptable limits.

\begin{figure}[ht]
\centering
\includegraphics [width=1\linewidth]{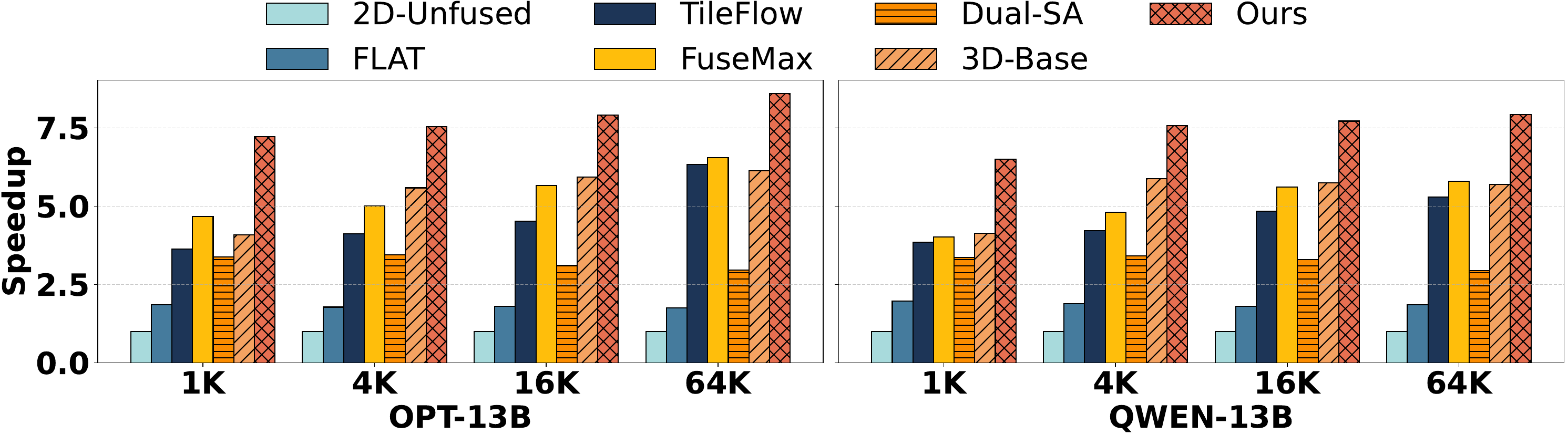}
\caption{Comparison of speedup of attention with sequence lengths ranging from \textit{1K} to \textit{64K}.}
\label{fig:speedup}
\end{figure}

\subsubsection{\textbf{Speedup and PE utilization}}  

\begin{figure}[ht]
\centering
\includegraphics [width=1.0\linewidth]{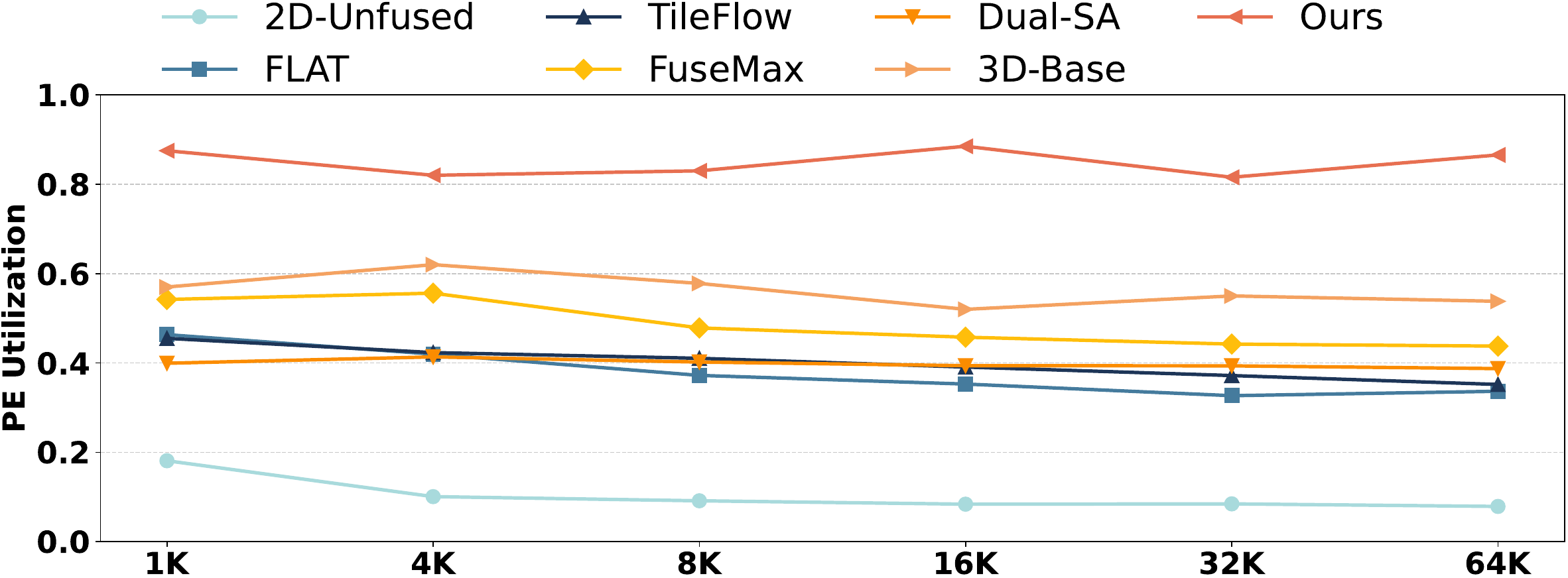}
\caption{Average utilization of PE arrays with sequence lengths ranging from \textit{1K} to \textit{64K}.}
\label{fig:utilization}
\end{figure}

Figure \ref{fig:speedup} shows inference speedup of various accelerators, normalized to the 2D-UnFused baseline. Compared to 2D-Unfused, FuseMax, Dual-SA, and 3D-Base baselines, our design achieves average speedups of 7.62$\times$, 1.46$\times$, 2.36$\times$, and 1.43$\times$, respectively. This improvement stems from our unique 3D architecture and dataflow co-design, which fundamentally eliminates the latency bottlenecks present in other architectures. Specifically, we map different operators of attention onto stacked PE array tiers, enabling intermediate results to transfer directly via low-latency inter-tier registers. Furthermore, softmax operations are performed within the PE array, avoiding offloading overhead and preventing a separate unit from becoming a system-wide bottleneck. In contrast, while Dual-SA attempts to bypass SRAM by using two PE arrays for $QK^T$and $PV$, 
but sparse and high-latency inter-array interconnects in its 2D layout create significant ``drain-and-inject" overhead, and its separate Softmax unit further obstructs the critical path. This high performance is complemented by excellent hardware efficiency: as shown in Figure~\ref{fig:utilization}, our 3D architecture achieves an average PE utilization of 87\% by reducing DRAM and SRAM accesses to prevent PE stalls and employing a carefully designed inter-tier pipeline for the entire attention computation across vertical tiers, minimizing pipeline bubbles.

%% file: Text/6.Conclusion.tex
\section{CONCLUSION}

This paper proposes 3D-Flow, a hybrid-bonded, 3D-stacked spatial accelerator that enables register-to-register communication across vertically stacked PE tiers. Unlike 2D multi-array architectures limited by NoC-based router-to-router transfers, 3D-Flow leverages sub-10 µm vertical TSVs to sustain cycle-level operator pipelining with minimal overhead. On top of this architecture, we design 3D-FlashAttention, a fine-grained scheduling method that balances latency across tiers, forming a bubble-free vertical dataflow without on-chip SRAM round-trips. Evaluations on Transformer workloads show that our 3D spatial accelerator reduces 46–93\% energy consumption and achieves 1.4×–7.6× speedups compared to state-of-the-art 2D and 3D designs.